\documentclass[a4paper]{article}
\usepackage{Odyssey2020}
\usepackage{epsfig,amssymb,amsmath}
\ninept
\usepackage{multirow}
\usepackage{subfigure}
\usepackage{mathrsfs}
\usepackage{float}
\usepackage{hyperref}
\usepackage{url}
\usepackage[ruled,vlined]{algorithm2e}
\usepackage{cite}
\title{Transforming Spectrum and Prosody for Emotional Voice Conversion with Non-Parallel Training Data}

\name{Kun Zhou$^{1}$,\thanks{\textbf{Codes \& Speech Samples:} \url{https://kunzhou9646.github.io/Odyssey2020_emotional_VC//}} Berrak Sisman$^{1,2}$, Haizhou Li$^{1}$}

\address{$^{1}$ Dept. of Electrical and Computer Engineering, National University of Singapore, Singapore \\ $^{2}$ Information Systems Technology and Design, Singapore University of Technology and Design, Singapore \\
{\small \tt zhoukun@u.nus.edu, berraksisman@u.nus.edu, haizhou.li@nus.edu.sg} }
\begin{document}
\maketitle

\begin{abstract}
Emotional voice conversion aims to convert the spectrum and prosody to change the emotional patterns of speech, while preserving the speaker identity and linguistic content.
Many studies require parallel speech data between different emotional patterns, which is not practical in real life. Moreover, they often model the conversion of fundamental frequency (F0) with a simple linear transform. As F0 is a key aspect of intonation that is hierarchical in nature, we believe that it is more adequate
to model F0 in different temporal scales by using wavelet
transform. 
%We propose to use CycleGAN for emotional voice conversion with non-parallel training data. We study novel training strategies for spectrum and prosody mapping models. 
%By learningforward and inverse mappings simultaneously using adversarial and cycle-consistency losses, 
We propose a CycleGAN network to find an optimal
pseudo pair from non-parallel training data by learning
forward and inverse mappings simultaneously using adversarial
and cycle-consistency losses. We also study the use of continuous wavelet transform (CWT) to decompose
F0 into ten temporal scales that describes speech prosody at different time resolution, for effective F0 conversion. Experimental results show that our
proposed framework outperforms the baselines both in objective and subjective evaluations. \\

%so far, many emotional VC frameworks work under an assumption of parallel training data, which is not practical in real life. Moreover, they often model fundamental frequency (F0) with a simple representation and transform it using a linear method. We note that  F0 is a key aspect of intonation and it is hierarchical in nature. Therefore, we believe that it is more adequate to model F0 in different temporal scales by using wavelet transform. \\
\end{abstract}
\section{Introduction}
Emotion, as an essential component of human communication, can be conveyed by various prosodic features, such as pitch, intensity, and speaking rate \cite{scherer1991vocal}. It plays an important role as a manifestation at semantic and pragmatic level of spoken languages. An adequate rendering of emotion in speech is critically important in expressive text-to-speech \cite{liu2020wavetts,liu2019teacher}, personalized speech synthesis, and intelligent dialogue systems, such as social robots and conversational agents. 

\begin{figure}[t]
    \centering
    \includegraphics[scale=0.6]{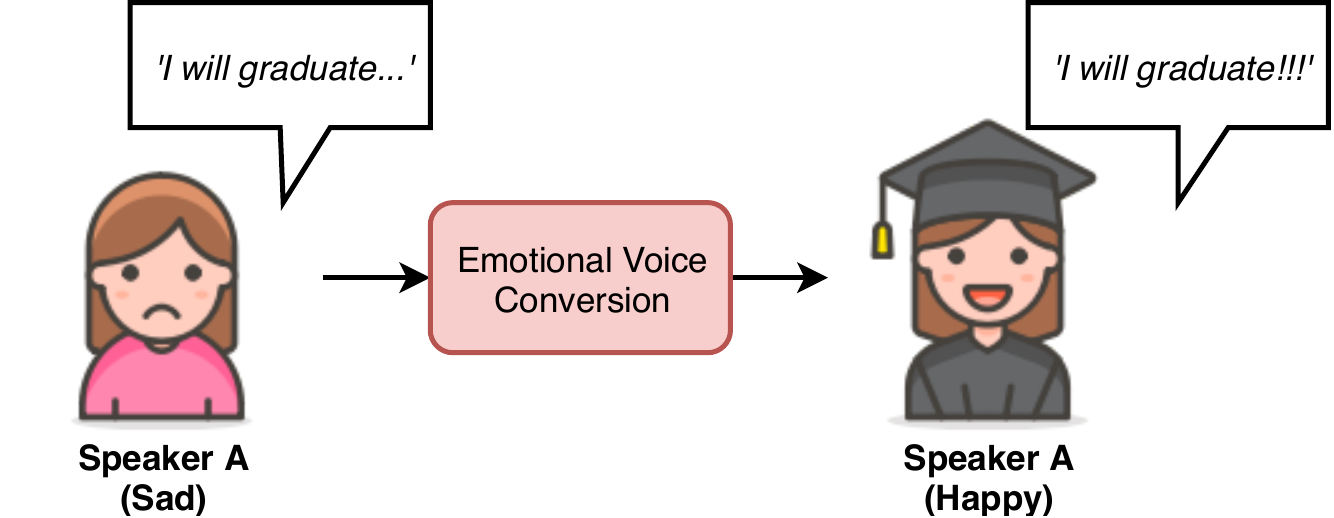}
    \vspace{-6mm}
    \caption{An emotional voice conversion system is trained on speech data of different emotional patterns from the same speaker. At run-time, the system takes the speech of one emotion as input, and converts to that of another \cite{aihara2012gmm,aihara2014exemplar,gao2018nonparallel}.}
    \label{fig:em-conversion}
    \vspace{-6mm}
\end{figure}

Emotional voice conversion is a voice conversion (VC) technique for converting the emotion from the source utterance to the target utterance, while preserving the linguistic information and the speaker identity, as illustrated in Figure \ref{fig:em-conversion}. It shares many similarities with conventional voice conversion. Both of them aim to convert non-linguistic information through mapping features from source to target. They are also different because conventional voice conversion techniques consider prosody-related features as speaker-independent. As speaker identity is thought to be characterized by the physical attributes of the speaker, which are strongly affected by the spectrum and determined by the voice quality of the individual \cite{ramakrishnan2012speech}, conventional VC studies mainly focus on spectrum conversion. On the other hand, emotion is inherently supra-segmental and hierarchical in nature \cite{xu2011speech,latorre2008multilevel}, that is manifested both in spectrum and prosody. Therefore, emotion cannot be handled simply at frame level, as it is insufficient to just convert the spectral features frame-by-frame.

Early studies of VC marked a success by training the spectral mapping on parallel speech data between source and target speaker \cite{abe1990voice,shikano1991speaker}. Many statistical approaches have been proposed in the past decades, such as Gaussian Mixture Model (GMM) \cite{toda2007voice}, and Partial Least Square Regression (PLSR) \cite{helander2010voice}. Other VC methods, such as Non-negative Matrix Factorization (NMF) \cite{lee2001algorithms} and exemplar-based sparse representation schemes \cite{wu2014exemplar,ccicsman2017sparse,sisman2018voice} were designed to address the over-smoothing problem in VC.

With the advent of deep learning, the performance of VC systems has been markedly improved. Neural Network (NN)-based methods, such as Restricted Boltzmann Machine (RBM) \cite{chen2014voice}, Feed Forward NN \cite{desai2010spectral}, Deep Neural Network (DNN) \cite{hinton2006fast}, and Recurrent Neural Network (RNN) \cite{nakashika2014high} have helped VC systems to achieve a higher level in terms of modeling the relationship between source and target features. More recently, some approaches have been proposed to eliminate the need of parallel data for VC such as Deep Bidirectional Long-Short-Term Memory (DBLSTM) with i-vector \cite{wu2016use}, variational auto-encoder \cite{hsu2016voice}, DBLSTM with model using Phonetic Posteriorgrams (PPGs) \cite{sun2016phonetic}, and GANs  \cite{hsu2017voice, kaneko2017parallel,kaneko2018cyclegan,sismanstudy}. The successful practice of these deep learning methods became the source of inspiration for this study. 

 The early studies on emotional VC \cite{tao2006prosody,wu2009hierarchical} only focused on prosody conversion by using a classification and regression tree to decompose the pitch contour of the source speech into a hierarchical structure, then followed by GMM and regression-based clustering methods. One attempt to handle both spectrum and prosody conversion \cite{schroder2001emotional,iida2003corpus,an2017emotional} was the GMM-based technique \cite{aihara2012gmm}. Another approach is a combination of Hidden Markov Model (HMM), GMM and F0 segment selection method for transforming F0, duration and spectrum, which was proposed in \cite{inanoglu2009data}. More recently, exemplar-based emotional VC approach based on NMF \cite{aihara2014exemplar} and other NN-based models such as DNN \cite{lorenzo2018investigating}, Deep Belief Network (DBN) \cite{luo2016emotional} and DBLSTM \cite{ming2016deep} were also proposed to perform spectrum and prosody mapping. Inspired by the success of sequence-to-sequence models in text-to-speech synthesis, a sequence-to-sequence encoder-decoder based model \cite{robinson2019sequence} was also investigated to transform the intonation of a human voice, and can convert the emotion of neutral utterances effectively. Rule-based emotional VC approaches such as \cite{xue2018voice} are capable of controlling the degree of emotion using dimensional space such as arousal and valence.
 
 We note that the training of most of the emotional VC systems relies on parallel training data, which is not practical in real life applications. Motivated by that, more recently, a style transfer auto-encoder \cite{gao2018nonparallel} was proposed, which can learn from non-parallel training data. A source-target pair in non-parallel dataset represents source and target emotions. But, unlike those in parallel dataset, they can carry different linguistic content, that may make data collection much easier.

Prosody conveys linguistic, para-linguistic and various types of non-linguistic information, such as speaker identity, emotion, intention, attitude and mood. It is observed that prosody is influenced by short-term as well as long-term dependencies \cite{sanchez2014hierarchical, sisman2019group}. We note that F0 is an essential prosodic factor with respect to the intonation in speech, describing the variation of the vocal pitch over different time domains, from the syllables to the entire utterance. Therefore, it should be represented with hierarchical modeling \cite{ming2015fundamental, ming2016exemplar,csicsman2017transformation}, for example, in multiple time scales. The early studies on emotional voice conversion use a Logarithm Gaussian (LG)-based linear transformation method \cite{tao2006prosody,wu2009hierarchical,aihara2012gmm,aihara2014exemplar,gao2018nonparallel} to convert F0. Such single pitch value of F0 representation does not characterize speech prosody well \cite{xu2011speech, latorre2008multilevel, sisman2019group}. Continuous Wavelet Transform (CWT) decomposes a signal into frequency components and represent it with different temporal scales, that becomes an excellent instrument. CWT has already been applied for speech voice conversion frameworks such as DKPLS \cite{sanchez2014hierarchical} and exemplar-based conversion \cite{sisman2018wavelet,csicsman2017transformation}. It has been also shown to be effective for emotional voice conversion such as NMF-based approach \cite{ming2016exemplar, ming2015fundamental} and DBLSTM-based approach \cite{ming2016deep}; and for emotional speech synthesis have been investigated in \cite{luo2017emotional,luo2017emotion,luo2019emotional}. 

%In this paper, we will be using CWT to model F0 in different temporal scales, and perform feature conversion with CycleGAN to achieve parallel-data-free emotional voice conversion.

\begin{figure*}
\centering
\subfigure['It is well never to know an author' in a neutral tone.]{
\begin{minipage}[c]{0.5\linewidth}
\centering
\includegraphics[width=7.9cm]{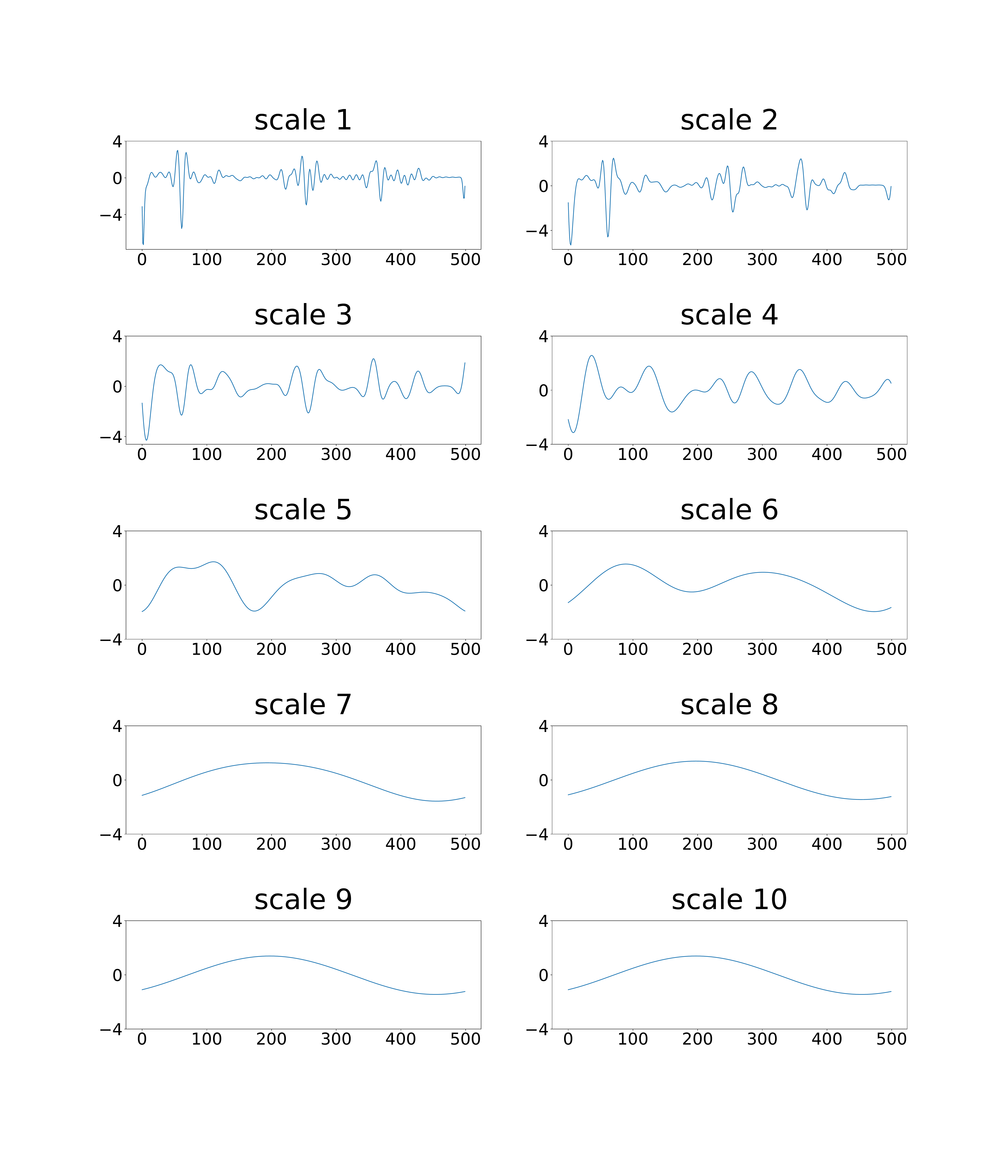}
\vspace{-6mm}
%\caption{10-scales CWT of F0 for a neutral voice}
\end{minipage}%
}%
\subfigure['It is well never to know an author' in an angry tone.]{
\begin{minipage}[c]{0.5\linewidth}
\centering
\includegraphics[width=7.9cm]{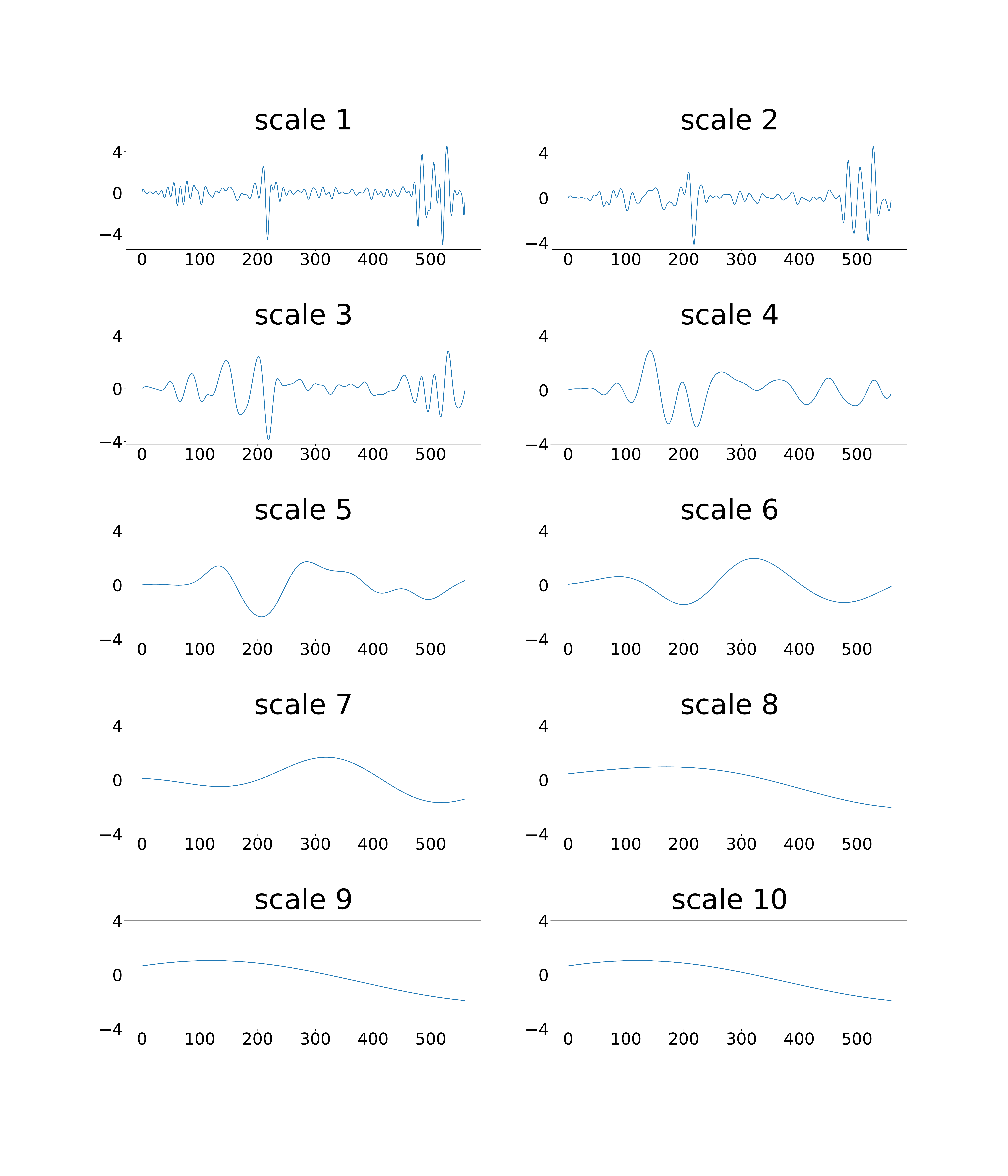}
\vspace{-6mm}
%\caption{10-scales CWT of F0 for an angry voice}
\end{minipage}%
}%
\centering
    \vspace{-4mm}
\caption{10-scales CWT analysis of F0 \cite{csicsman2017transformation,luo2019emotional} of an utterance in neutral and angry tone with the same linguistic content.}
\vspace{-3mm}
\label{fig:cwt}
\end{figure*}
In this paper, we propose an emotional VC framework with CycleGAN that is trained on non-parallel data to map a speaker's speech from one emotion to another. We use mel-spectrum to represent the acoustic features and CWT coefficients for prosodic features. Our framework does not rely on either parallel training data or any other extra modules such as speech recognition or time alignment procedures. 

The main contributions of this paper include: 1) we propose a parallel-data-free emotional voice conversion framework; 2) we show the effect of prosody for emotional voice conversion; 3)  we effectively convert spectral and prosodic features with CycleGAN; 4) we investigate different training strategies for spectrum and prosody conversion such as separate training and joint training; and 5) we outperform the baseline approaches, and achieve quality converted voice.  

% we compared our framework with the conventional CycleGAN-based VC framework, which uses CycleGAN to convert the spectrum and use LG-based linear transformation method to convert F0. 3) we investigated different training strategies, such as separate training and joint training of our framework, and reported the results.  \\

This paper is organized as follows: In Section 2, we describe the details of CycleGAN and CWT decomposition of F0. In Section 3, we explain our proposed spectrum and prosody conversion for emotional VC framework. Section 4 reports the experimental results. Conclusion is given in Section 5.

\section{Related Work}
\subsection{CycleGAN}
Recently, generative adversarial learning has become very popular in machine learning applications, such as computer vision \cite{emir2019semantically,ak2020semantically, zhang2017stackgan, ak2019attribute} and speech information processing \cite{berrak_ganslt, singan-2019}. In this paper, we focus on a GAN-based network called CycleGAN, which is capable of learning a mapping between source $x\in X$ and target $y\in Y$ from non-parallel training data. It is based on the concept of adversarial learning \cite{goodfellow2014generative}, which is to train a generative model to find a solution in a min-max game between two neural networks, called as generator ($G$) and discriminator ($D$). CycleGAN was first proposed for computer vision \cite{zhu2017unpaired, lu2017conditional}, and then extended to various fields including speech synthesis and voice conversion \cite{kaneko2017parallel,kaneko2018cyclegan,kaneko2019cyclegan}.

A CycleGAN is incorporated with three losses: adversarial loss, cycle-consistency loss, and identity-mapping loss, learning forward and inverse mapping between source and target. Adversarial loss measures how distinguishable between the data distribution of converted data and source or target data. For the forward mapping, it is defined as:
\begin{multline}
   L_{ADV}(G_{X\rightarrow Y},D_Y,X,Y) = \mathbb{E}_{y\sim P(y)}[D_Y(y)] \\
    +\mathbb{E}_{x\sim P(x)}[\log(1-D_Y(G_{X\rightarrow Y}(x))]
\end{multline}
The closer the distribution of converted data with that of target data, the smaller $L_{ADV}$ becomes. The adversarial loss only tells us whether $G_{X\rightarrow Y}$ follows the distribution of target data but does not help to preserve the contextual information. In order to guarantee that the contextual information of $x$ and $G_{X\rightarrow Y}$ will be consistent, the cycle-consistency loss is given as:
\begin{multline}
    L_{CYC}(G_{X\rightarrow Y},G_{Y\rightarrow X}) \\
    = \mathbb{E}_{x\sim P(x)}[\Vert G_{Y\rightarrow X}(G_{X\rightarrow Y}(x))-x\Vert_1]\\
    +\mathbb{E}_{y\sim P(y)}[\Vert G_{X\rightarrow Y}(G_{Y\rightarrow X}(y))-y\Vert_1]
\end{multline}
This loss encourages $G_{X\rightarrow Y}$ and $G_{Y\rightarrow X}$ to find an optimal pseudo pair of $(x,y)$ through circular conversion. To preserve the linguistic information without any external processes, an identity mapping loss is introduced as below:
\begin{multline}
    L_{ID}(G_{X\rightarrow Y},G_{Y\rightarrow X}) \\
    = \mathbb{E}_{x\sim P(x)}[\Vert G_{Y\rightarrow X}(x)-x\Vert]
    + \mathbb{E}_{y\sim P(y)}[\Vert G_{X\rightarrow Y}(y)-y\Vert]
\end{multline}
%Thus, full loss function is defined as: 
%\begin{multline}
 %   L(G,F,D_X,D_Y,X,Y) = L_{GAN}(G,D_Y,X,Y)\\
  %  +L_{GAN}(F,D_X,X,Y)+\lambda_{CYC}L_{CYC}(G,F,X,Y)\\
  %  +\lambda_{ID}L_{ID}(G,F,X,Y)
%\end{multline}
%where $\lambda_{CYC}$ and $\lambda_{ID}$ are trade-off parameters. \\

%The optimal mapping functions $G^*$ and $F^*$ are obtained by solving the minmax-game defined as:
%\begin{equation}
 %   G^*,F^*=\mathop{\arg\min}_{G,F}\mathop{\max}_{D_X,D_Y}L(G,F,D_X,D_Y,X,Y)
%\end{equation}

 We note that CycleGAN is well-known for achieving remarkable results without parallel training data in many fields from computer vision to speech information processing. In this paper, we propose to use CycleGAN for spectrum and prosody conversion for emotional voice conversion with non-parallel training data. 
 
\subsection{Continuous Wavelet Transform (CWT)}
It is well-known that emotion can be conveyed by various prosodic features, such as pitch, intensity and speaking rate. F0 is an essential part with respect to the intonation. We note that the modeling of F0 is a challenging task as F0 is discontinuous due to the unvoiced parts, and hierarchical in nature.  As a multi-scale modeling method, CWT makes it possible to decompose F0 to different variations over multiple time scales. 

Wavelet transform provides an easily interpretable visual representation of signals. Using CWT, a signal can be decomposed into different temporal scales. We note that CWT has been successfully used in speech synthesis \cite{kruschke2003estimation,mishra2006decomposition} and voice conversion \cite{sisman2019group, sisman2018wavelet}.

\begin{figure*}
\vspace{-3mm}
    \centering
    \includegraphics[width=17cm]{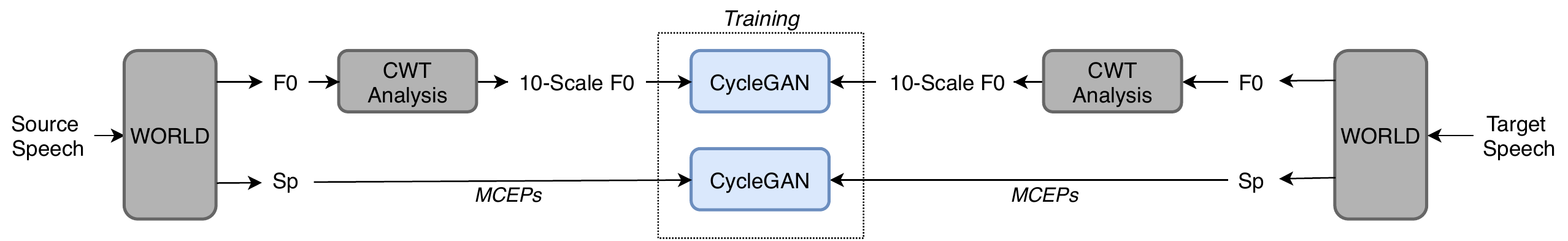}
    \vspace{-6mm}
    \caption{The training phase of the proposed CycleGAN-based emotional VC framework with WORLD vocoder. CWT is used to decompose F0 into 10 scales. Blue boxes are involved in the training, while grey boxes are not.}
    \label{fig:training}
\end{figure*}
\begin{figure*}
\vspace{-4mm}
    \centering
    \includegraphics[width=17cm]{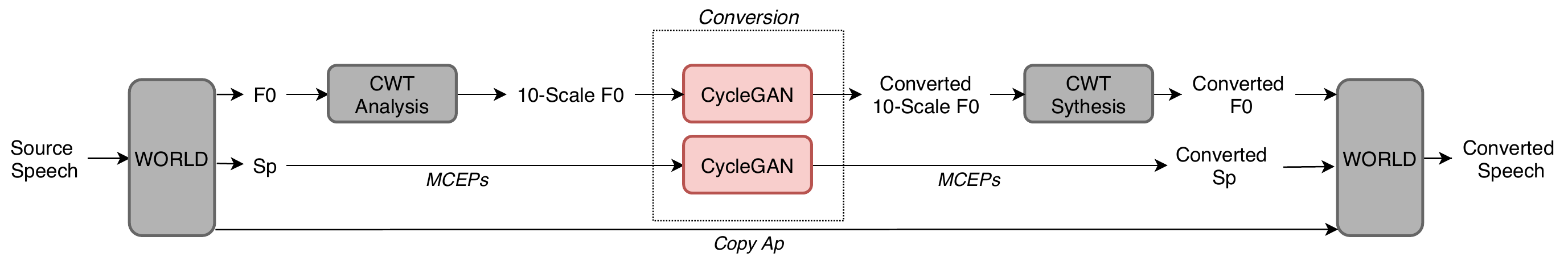}
    \vspace{-6mm}
    \caption{The run-time conversion phase of the proposed CycleGAN-based emotional VC framework. Pink boxes represent the networks which have been trained during the training phase.}
    \label{fig:conversion}
\end{figure*}

Given a bounded, continuous signal $k_0$, its CWT representation $W(k_0)(\tau,t)$ can be written as:
\begin{equation}
    W(k_0)(\tau,t)=\tau^{-1/2}\int_{-\infty}^{+\infty}k_0(x)\psi(\frac{x-t}{\tau})dx
\end{equation}
where $\psi$ is the Mexican hat mother wavelet. The original signal $k_0$ can be recovered from the wavelet representation $W(k_0)$ by inverse transform, given as:
\begin{equation}
k_0(t)=\int_{-\infty}^{+\infty}\int_{0}^{+\infty}W(k_0)(\tau,x)\tau^{-5/2}\psi(\frac{t-x}{\tau})dxd\tau    
\end{equation}
However, if all information on $W(k_0)$ is not available, the reconstruction is incomplete. In this study, we fix the analysis at ten discrete scales, one octave apart. The decomposition is given as:
\begin{equation}
    W_i(k_0)(t)=W_i(k_0)(2^{i+1}\tau_0,t)(i+2.5)^{-5/2}
\end{equation}
The reconstructed $k_0$ is approximated as:
\begin{equation}
    k_0(t) = \sum_{i=1}^{10}W_i(k_0)(t)(i+2.5)^{-5/2}
\label{eq:restruct}
\end{equation}
where $i=1,...,10$ and $\tau = 5 ms$. These timing scales were originally proposed in \cite{suni2013wavelets} and in prosody model \cite{vainio2013continuous, sisman2018phonetically}. We believe that the prosody of emotion is expressed differently at different time scales.  With the multi-scale representations, lower scales capture the short-term variations and higher scales capture the long-term variations.  In this way, we are able to model and transfer the F0 variants from the micro-prosody level to the whole utterance level for emotion pairs. In  Figure \ref{fig:cwt}, we use an example to compare two utterances with the same content but different emotion across time scales. 

\section{Spectrum and Prosody Conversion for Emotional Voice Conversion}

In this section, we propose an emotional VC framework that performs both spectrum and prosody conversion using Cycle-Consistent Adversarial Networks. As an essential component of prosody, we propose to use CWT to decompose one-dimensional F0 into 10 time-scales. The proposed framework is trained on non-parallel speech data, eliminating the need of parallel training data and effectively converts the emotion of source speaker from one state to another.

The training phase of the proposed framework is given in Figure \ref{fig:training}. We first extract spectral and F0 features from both source and target utterances using WORLD vocoder \cite{morise2016world}. It is noted that F0 features extracted from WORLD vocoder are discontinuous, due to the voiced/unvoiced parts within an utterance. Since CWT is sensitive to the discontinuities in F0, we perform the following pre-processing steps for F0: 1) linear interpolation over unvoiced regions, 2) transformation of F0 from linear to logarithmic scale, and 3) normalization of the resulting F0 to zero mean and unit variance. We then perform the CWT decomposition of F0 as given in Eq. (6) and Algorithm 1. 

We train CycleGAN for spectrum conversion with 24-dimensional Mel-cepstral coefficients (MCEPs), and for prosody conversion with 10-dimensional F0 features for each speech frame. We note that the source and target training data are from the same speaker, but consist of different linguistic content and different emotions.  By learning forward and inverse mappings simultaneously using adversarial and cycle-consistency losses, we encourage CycleGAN to find an optimal mapping between source and target spectrum and prosody features. 

The run-time conversion phase is shown in Figure \ref{fig:conversion}. We first use the WORLD vocoder to extract spectral features, F0, and aperiodicity (AP) from a given source utterance. Similar to that of training phase,
we encode spectral features as 24-dimensional MCEPs, and obtain 10-scale F0 features through CWT decomposition of F0, that is also reported in Algorithm 1. 24-dimensional MCEPs and 10-scale F0 are fed into the corresponding trained CycleGAN models to perform spectrum and prosody conversion separately. We reconstruct the converted F0 with CWT synthesis approximation method, that is given in Eq. (\ref{eq:restruct}) and Algorithm 2. Finally, we use WORLD vocoder to synthesize the converted emotional speech. 
\vspace{-3mm}
\begin{algorithm}[h]
\label{algo:main}
%\setstretch{1.1}
\SetAlgoLined
% \KwResult{Write here the result }
\textbf{Input}:\\
 \quad The signal, $k_0[m], m = 0,1,...,N-1$\\
 \quad Mother wavelet, $\psi(t)$\\
 \quad Scale, $i$\\
 \quad $T$, where $[-T,T]$ is the support of $\psi(t)$\\

\textbf{Output}:\\
 \quad Wavelet transform: $W_i(k_0)[n], n = 0,1,...,N-1$

\textbf{Parameter}:\\
 \quad $\tau$: sampling interval, set to be 0.005 \\
 \quad $d_j$: space between discrete scales, set to be 0.5  \\
 \quad $s_0$: starting scale, is set to be $2*\tau$ \\

\textbf{Begin}
%\\1. Sampling $f_0(t)$, then get the signal $f_0[m], m = 0,1,...,N-1$;
\\1. Let $\tilde{k} = [k_0,k_1,...,k_{N-1},\underbrace{0,...,0}_{2iT}]^T$;
\\2. Let $\tilde{\psi}_i(t) = \frac{1}{i}\Bar{\psi(\frac{iT-t}{i})}$;
\\3. Let $h_i[n] = \tilde{\psi}_i(m), m = 0,1,...,2iT$;
\\4. Let $\tilde{h}_i = [h_i[0], h_i[1],...,h_i[2iT],\underbrace{0,...,0}_{N-1}]^T$;
\\5. $W_i[n] = ifft(fft(\tilde{k}) * fft(\tilde{h}_i))$, \\\quad$n = 0, 1,..., 2iT + N - 1$;
\\6. $W_i(k_0)[n] = W_i[n+iT], n = 0,1,...,N-1$;
\\\textbf{Return}: $W_i(k_0)[n], n = 0,1,...,N-1$
\\\textbf{End}
 \caption{CWT Decomposition}
\end{algorithm}

\vspace{-5mm}
\section{Experiments}
We conduct both objective and subjective experiments to assess the performance of our proposed parallel-data-free emotional VC framework. In this paper, we use the emotional speech corpus \cite{liu2014emotional}, which is recorded by a professional American actress, speaking English utterances with the same content in seven different emotions. We randomly choose four emotions, that are 1) neutral, 2) angry, 3) sad, and 4) surprise. 

We perform CWT to decompose F0 into 10 different scales and train CycleGAN using non-parallel training data to learn the relationships of spectral and prosody features between different emotions of the same speaker. CycleGAN-based spectrum conversion framework, denoted as \textit{baseline}, is used as the reference framework. In this framework, F0 is transformed through LG-based linear transformation method. 

We are also interested in the effect of joint and separate training for spectrum and prosody features. In joint training, we concatenate 24 MCEPs and 10 CWT coefficients to form a vector for each frame to train the joint spectrum-prosody CycleGAN. In separate training, we train a spectrum CycleGAN with the MCEP features, and a prosody CycleGAN with the CWT coefficients separately. Hereafter, we denote the separate training as \textit{CycleGAN-Separate}, and the joint training as \textit{CycleGAN-Joint}. The comparison of the frameworks can be also seen in Table \ref{tab:compare}.

\begin{table*}[t]
\vspace{-10mm}
\centering
\begin{tabular}{|c|c|c|}
\hline
\textbf{Framework}         & \textbf{Spectrum Conversion}                & \textbf{Prosody Conversion (F0)}               \\ \hline \hline
\textit{Baseline}          & Spectrum CycleGAN  & LG-based F0 linear transformation   \\ \hline \hline
\textit{CycleGAN-Joint }   & \multicolumn{2}{c|}{Joint Spectrum-Prosody CycleGAN} \\ \hline \hline
\textit{CycleGAN-Separate (proposed)} & Spectrum CycleGAN & Prosody CycleGAN \\ \hline \hline

\end{tabular}
\vspace{-2mm}
\caption{The comparison of the baseline, CycleGAN-Joint, and CycleGAN-Separate for spectrum and prosody conversion.}
\label{tab:compare}
\vspace{-3mm}
\end{table*}

\subsection{Experimental Setup}
The speech data in \cite{liu2014emotional} is sampled at 16kHz with 16-bit per sample. The audio files for each emotion are manually segmented into 100 short parallel sentences (approximately 3 minutes). Among them, 90 and 10 sentences are provided as training and evaluation sets, respectively. In order to make sure that our proposed model is trained under non-parallel condition, the first 45 utterances are used for the source and the other 45 sentences are used for the target. 24 Mel-cepstral coefficients (MCEPs), fundamental frequency (F0), and APs are then extracted every 5 ms using WORLD vocoder \cite{morise2016world}. As a pre-processing step, we normalize the source and target MCEPs per dimension. 

We report the performance of three frameworks that use CycleGAN, namely 1) baseline 2) CycleGAN-Joint, and 3) CycleGAN-Separate. For the baseline, we extract 24-dimensional MCEPs and one-dimensional F0 features for each frame. For both CycleGAN-Separate and CycleGAN-Joint, each speech frame is represented with 24-dimensional MCEPs and 10-dimensional F0 features. We adopt the same network structure for all frameworks. We design the generators using a one-dimensional (1D) CNN to capture the relationship among the overall features while preserving the temporal structure. The 1D CNN is incorporated with down-sampling, residual, and up-sampling layers. As for the discriminator, a 2D CNN is employed. For all frameworks, we set $\lambda_{CYC} = 10$. $L_{ID}$ is only used for the first $10^4$ iterations with $\lambda_{ID}=5$ to guide the learning process. 

We train the networks using the Adam optimizer with a batch size of 1. We set the initial learning rates to 0.0002 for the generators and 0.0001 for the discriminators. We keep the learning rate the same for the first $2\times 10^5$ iterations, which then linearly decays over the next $2\times 10^5$ iterations. The momentum term $\beta_1$ is set to be 0.5. As CycleGAN does not require source-target pair to be the same length, time alignment is not necessary.

\subsection{Objective Evaluation}
We perform objective evaluation to assess the performance of both spectrum and prosody conversion. In all experiments, we use 45-45 non-parallel utterances during training. 
\subsubsection{Spectrum Conversion}
We employ Mel-cepstral distortion (MCD) between the converted and target Mel-cepstra  to measure the spectrum conversion, that is given as follows:
\begin{equation}
    MCD[dB]=(10/\ln{10})\sqrt{2\sum_{i=1}^{24}(mceps_i^{t}-mceps_i^{c})^2}
\end{equation}
where $mceps_i^{c}$ and $mceps_i^{t}$ represent the converted and target MCEPs sequences, respectively. A lower MCD indicates better performance.

\begin{table}[h]
\vspace{-1mm}
\centering
\begin{tabular}{|c|c|c|c|c|}
\hline
\multirow{2}{*}{} & \multicolumn{2}{c|}{MCD {[}dB{]}}  \\ \cline{2-3} 
                  & CycleGAN-Joint & CycleGAN-Separate \\ \hline
Neutral$\rightarrow$Angry     & 10.87          & 8.83              \\ \hline
Neutral$\rightarrow$Sad       & 9.41           & 8.27              \\ \hline
Neutral$\rightarrow$Surprise  & 10.43          & 9.05              \\ \hline
Overall mean           & 10.23          & \textbf{8.71}              \\ \hline
\end{tabular}
\vspace{-2mm}
\caption{A comparison of the MCD results between CycleGAN-Joint and CycleGAN-Separate for three different emotion combinations. }
\label{tab:table1}
\vspace{-1mm}
\end{table}

Table \ref{tab:table1} reports the MCD values for a number of settings in a comparative study. The MCD values are calculated for both joint and separate training of spectrum and prosody features. We conducted the experiments for three emotion combinations: 1) neutral-to-angry, 2) neutral-to-sad, and 3) neutral-to-surprise. We observed that all separate training settings consistently outperform those of joint training settings by achieving lower MCD values. For example, the overall MCD of separate training is 8.71, while it is 10.23 for joint training. 

We note that the baseline trains CycleGAN only with spectral features. Therefore, its spectral distortion is supposed to be the same with that of CycleGAN-Separate. That is the reason why MCD results of the baseline do not need to report in this case.

%%%%%%%%%%%%%%%%%%%%%%%%%%%%%%%%%%%%%%%%%%%%%%%%%%%%%%%

\subsubsection{Prosody Conversion}
We use Pearson Correlation Coefficient (PCC) and Root Mean Squared Error (RMSE) to report the performance of prosody conversion \cite{sisman2019group}. The RMSE between the converted F0 and the corresponding target F0 is defined as:
\begin{equation}
    RMSE=\sqrt{\frac{1}{N}\sum_{i=1}^{N}(F0_{i}^c-F0_{i}^t)^2}
\end{equation}
where $F0_{i}^c$ and $F0_{i}^t$ denote the converted and target interpolated F0 features, respectively. $N$ is the length of $F0$ sequence. We note that a lower RMSE value represents better F0 conversion performance. 

The PCC between the converted and target F0 sequences is given as:
\begin{equation}
    \rho(F0^c,F0^t) = \frac{cov(F0^c,F0^t)}{\sigma_{F0^c}\sigma_{F0^t}} 
\end{equation}
where $\sigma_{F0^c}$ and $\sigma_{F0^t}$ are the standard deviations of the converted F0 sequences ($F0^c$) and the target F0 sequences ($F0^t$), respectively. We note that a higher PCC value represents better F0 conversion performance. 

\begin{table*}[t]
\vspace{-12mm}
\begin{tabular}{|c|c|c|c|c|c|c|}
\hline
\multirow{2}{*}{} & \multicolumn{3}{c|}{RMSE {[}Hz{]}}            & \multicolumn{3}{c|}{PCC}                     \\ \cline{2-7} 
                  & Baseline & CycleGAN-Joint & CycleGAN-Separate & Baseline & CycleGAN-Joint & CycleGAN-Separate \\ \hline  \hline
Neutral$\rightarrow$Angry     & 71.09    & 64.55          & 67.44             & 0.75     & 0.81           & 0.78              \\ \hline \hline
Neutral$\rightarrow$Sad       & 62.99    & 57.46          & 48.33             & 0.66     & 0.68           & 0.74              \\ \hline \hline
Neutral$\rightarrow$Surprise  & 77.89    & 73.16          & 74.14             & 0.75     & 0.79           & 0.76              \\ \hline \hline
Overall mean      & 70.62    & 65.05          & \textbf{63.03}             & 0.72     & \textbf{0.76}           & \textbf{0.76}              \\ \hline \hline
\end{tabular}
\vspace{-3mm}
\caption{A comparison of the RMSE and PCC results of the baseline, CycleGAN-Joint and CycleGAN-Separate for three different emotion combinations (neutral-to-angry, neutral-to-sad and neutral-to-surprise). }
\label{tab:table2}
\vspace{-4mm}
\end{table*}

\begin{algorithm}[t]
\label{algo:main}
%\setstretch{1.1}
\SetAlgoLined
% \KwResult{Write here the result }
\textbf{Input}:\\
\quad Decomposed $k_0$ wavelet features:\\
\quad $W_i(k_0)[n], n = 0,1,...,N-1$\\
\quad Scale, $i$\\
\textbf{Output}:\\
\quad Reconstructed signal: $k_0[n], n = 0,1,...,N-1$\\
\textbf{Begin}
\\\textbf{for} i = 1, 2, ... ,10 :
\\\quad$k_i[n] = W_i(k_0)[n] * ((i+2.5)^{-2.5})$;
\\\quad$k_0[n] += k_i[n]$;
\\\textbf{end}
\\\textbf{Return}: $k_0[n],n = 0,1,...,N-1$
\\\textbf{End}
 \caption{CWT Synthesis}
\end{algorithm}

Table \ref{tab:table2} reports the RMSE and PCC values of F0 conversion for a number of settings in a comparative  study. In this experiment, we conducted three emotional conversion settings: 1) neutral-to-angry, 2) neutral-to-sad, 3) neutral-to-surprise. We also report the overall performance. As for RMSE results, first of all, we observe that the proposed prosody conversion, based on CycleGAN with CWT-based F0 decomposition outperforms  the traditional baseline (denoted as \textit{baseline}) where F0 is converted with LG-based linear transformation method. Secondly, the proposed separate training with CycleGAN for spectrum and CWT-based prosody conversion overall achieves better result (RMSE: 63.03) than separate training (RMSE: 65.05), which is also consistent with the objective evaluation. PCC results suggest that both joint and separate training of CWT-based F0 features achieve similar results. 

We would like to highlight that the proposed CWT-based modeling for F0 always outperforms the baseline framework that uses LG-based linear transformation method. 
\begin{figure}[t]
   \vspace{-3mm}
    \centering
    \includegraphics[width=7cm]{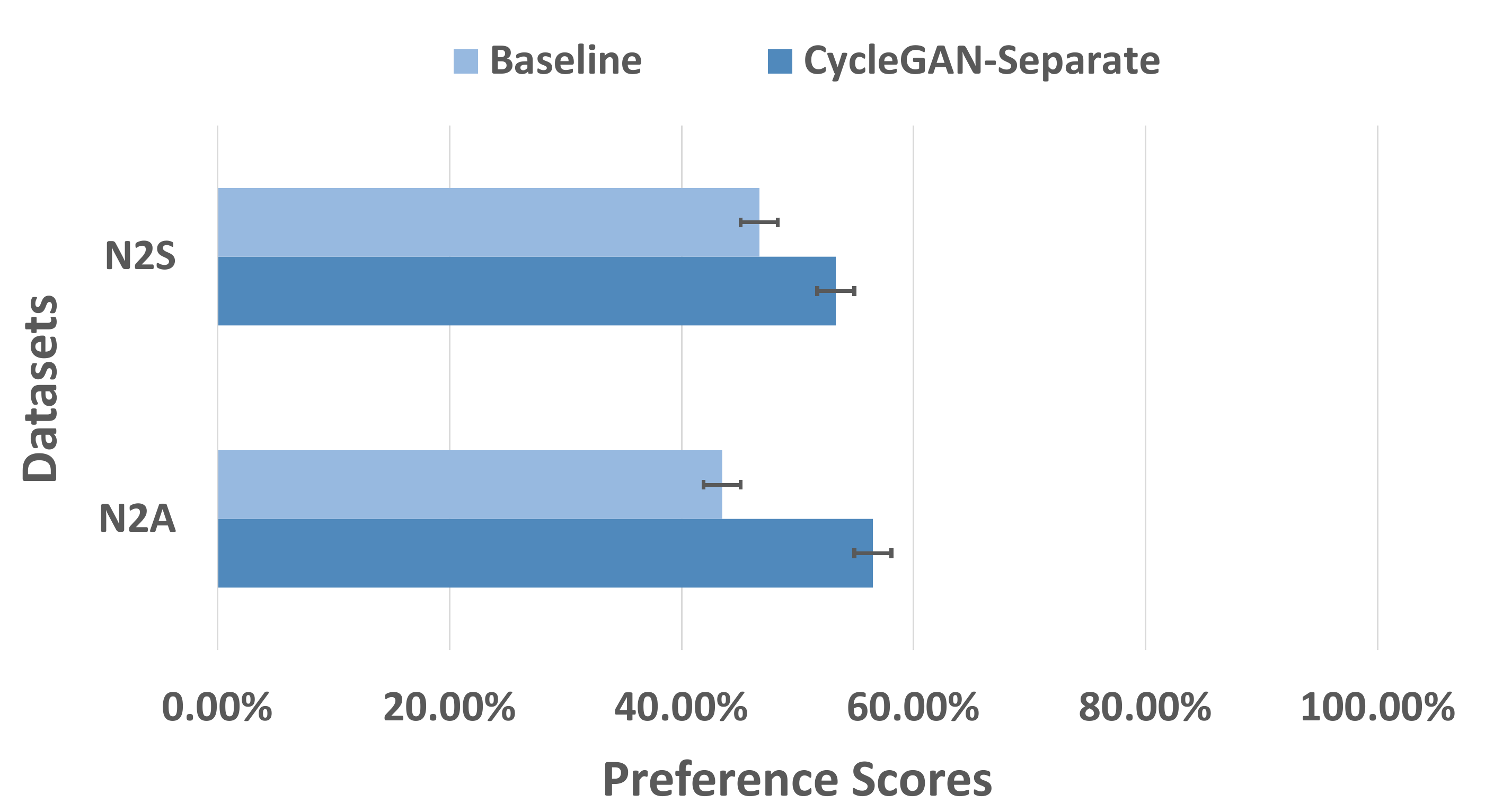}
    \vspace{-4mm}
    \caption{The XAB preference results with 95\% confidence interval between  the baseline and CycleGAN-Separate in emotion similarity experiments. }
    \label{fig:pref1}
    \vspace{-6mm}
\end{figure}

\subsection{Subjective Evaluation}
We further conduct two listening experiments to assess the proposed frameworks in terms of emotion similarity. We perform XAB test to assess the emotion similarity by asking listeners to choose the one which sounds more similar to the original target between A and B in terms of emotional expression. XAB test has been widely used in speech synthesis such as voice conversion \cite{sisman2019group}, singing voice conversion \cite{singan-2019} and emotional voice conversion \cite{luo2019emotional}. In both experiments, 45-45 non-parallel utterances are used during training. We selected two emotion combinations  for the listening experiments, that are 1) neutral-to-angry (N2A), and 2) neutral-to-surprise (N2S). 13 subjects participated in all the listening tests, each of them listens to 80 converted utterances in total.

We first conduct XAB test between the baseline and our proposed method to show the effect of our proposed framework that performs separate training of CycleGAN-based conversion for spectrum and CWT-based F0 modeling. Consistent with the previous experiments, our proposed framework is again denoted as \textit{CycleGAN-Separate}. Listeners are asked to listen to the source utterances, the baseline, our proposed method and the reference utterances respectively. Then, they are asked to choose the one which sounds more similar to the reference in terms of emotional expression. We note that both frameworks perform spectral conversion in the same way, while our proposed framework performs a more sophisticated F0 conversion, which is modeling with CWT, and then converting with CycleGAN. The results are reported in Figure \ref{fig:pref1} for 2 different emotional conversion scenarios that are N2A and N2S. We observe that the proposed CycleGAN-Separate outperforms the baseline framework in both experiments, which shows the effectiveness of prosody modeling and conversion, for emotional voice conversion.

\begin{figure}[t]
    \vspace{-3mm}
    \centering
    \includegraphics[width=7cm]{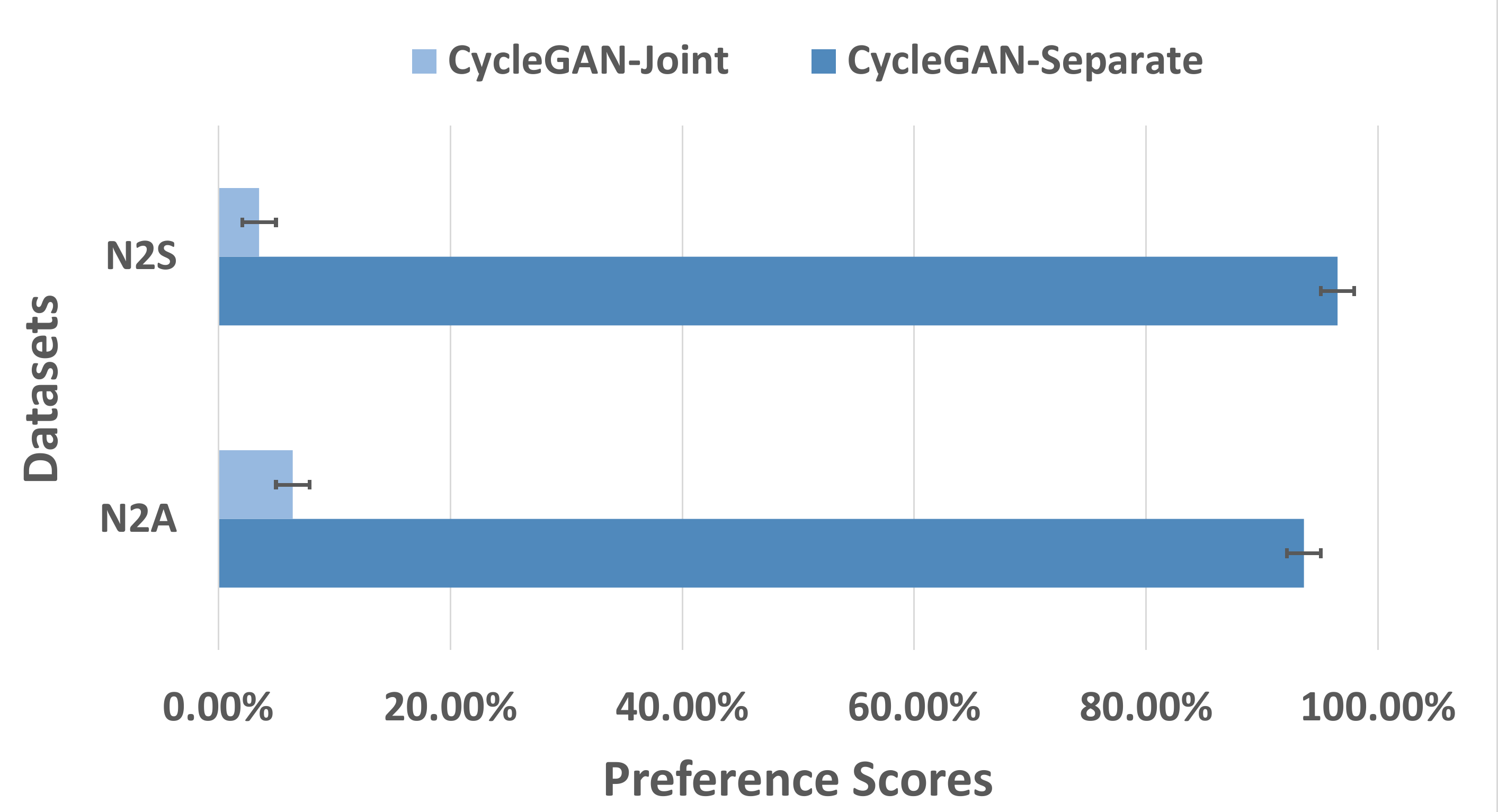}
    \vspace{-4mm}
    \caption{The XAB preference results with 95\% confidence interval between the CycleGAN-Joint and CycleGAN-Separate in emotion similarity experiments. }
    \label{fig:pref2}
   \vspace{-6mm}
\end{figure}

 We then conduct XAB test between joint and separate training to assess different training strategies for spectrum and prosody conversion. The results are reported in Figure \ref{fig:pref2} for two different emotional conversion scenarios N2A and N2S. We observed that the performance of separate training (denoted as \textit{CycleGAN-Separate}) is much better than the joint training (denoted as \textit{CycleGAN-Joint}). Our proposed method achieves 93.6$\%$ on N2A and 96.5$\%$ on N2S, which we believe are remarkable. 

%\section{Discussion}

\subsection{Joint vs. Separate Training of Spectrum and Prosody}
We observe that the listeners prefer the separate training much more than the joint training. We consider that prosody is manifested at different time scales, which also consists of content-dependent and content-independent elements.  

The joint training ties the CWT coefficients of F0 with the spectral features at the frame level, that assumes that prosody is content-dependent. With the limited number of training samples (45 pairs and around 3 minutes of speech), the CycleGAN model resulting from the joint training does not generalize well the emotional mapping for unseen content at run-time inference. With the separate training, the CycleGAN model is trained for spectrum and prosody separately. In this way, the prosody CycleGAN learns sufficiently well from the limited number of training samples between the emotion pairs in a content-independent manner. Therefore, separate training outperforms joint training in terms of emotion similarity.

\vspace{-1mm}
\section{Conclusion}
In this paper, we propose a high-quality parallel-data-free emotional voice conversion framework. We perform both spectrum and prosody conversion based on CycleGAN. We provide a non-linear method which uses CWT to decompose F0 into different timing-scales. Moreover, we also study the joint and separate training of CycleGAN for spectrum and prosody conversion. We observe that separate training of spectrum and prosody can achieve better performance than joint training, in terms of emotion similarity. Experimental results show that our proposed emotional voice conversion framework can achieve better performance than the baseline with non-parallel training data. 
\vspace{-3mm}
\section{Acknowledgments}
This work is supported by Human-Robot Interaction Phase 1 (Grant No. 192 25 00054), National Research Foundation Singapore under the National Robotics Programme. It is also supported by National Research Foundation Singapore under the AI Singapore Programme (Award Number: AISG-100E-2018-006), and Programmatic Grant No. A18A2b0046 (Human Robot Collaborative AI for AME) and A1687b0033 (Neuromorphic Computing) from the Singapore Government’s Research, Innovation and Enterprise 2020 plan in the Advanced Manufacturing and Engineering domain.

\bibliographystyle{IEEEbib}
{\footnotesize
\bibliography{Odyssey2020}}

% This could be also done as follows:
%
%\begin{thebibliography}{10}
%\bibitem[1]{aluisio2001learn}Sandra M. Alu\'{i}sio, Iris Barcelos, Jandir Sampaio, and Osvaldo
%N. Oliveira Jr, ``How to learn the many unwritten
%``rules of the game'' of the academic discourse: a hybrid
%approach based on critiques and cases to support scientific
%writing,'' in Proceedings of the IEEE International Conference
%on Advanced Learning Technologies, Madison, USA,
%August 2001, pp. 257–260.
%\bibitem[2]{swales1987writing} John Swales and Hazem Najjar, ``The writing of research
%article introductions,'' Written communication, vol. 4, no.
%2, pp. 175–191, 1987.
%\bibitem[3]{day2012write} Robert Day and Barbara Pastel, How to write and publish
%a scientific paper, Cambridge University Press, 2012.
%\bibitem[4]{teufel2000} Simone Teufel, Argumentative zoning: information extraction
%from scientific text, Ph.D. thesis, University of Edinburgh,
%2000.
%\bibitem[5]{berkenkotter1989social} Carol Berkenkotter, Thomas N. Huckin, and John Ackerman,
%``Social context and socially constructed texts: The
%initiation of a graduate student into a writing research community.
%technical report no. 33.,'' Tech. Rep., Center for
%the Study of Writing, University of California Berkeley \&
%Carnegie Mellon University, 1989.
%\end{thebibliography}

\end{document}